# Bonding and Electronic Properties of Ice at High Pressure

**B. Militzer**

*Department of Earth and Planetary Science and Department of Astronomy, University of California, Berkeley, USA*

*The properties of water ice at megabar pressure are characterized with ab initio computer simulations. The focus lies on the metallic Cmcm phase and its insulating distorted analogue with Pnma symmetry. Both phases were recently predicted to occur at 15.5 and 12.5 Mbar respectively [Phys. Rev. Lett. 105 (2010) 195701]. The Fermi surface of the Cmcm phase is analyzed and possibility of Fermi nesting to occur is discussed. The Wannier orbital are computed for the Pnma structure and compared to ice X. While ice X shows typical $sp^3$ hybridization, in the Pnma structure, the orbitals are deformed and no longer all aligned with the hydrogen bonds.*

## Introduction

Water is one of the most common binary compounds in the universe. In our solar system the majority of it occurs in the interiors of giant planets [1]. Uranus and Neptune consist of large amounts of water, ammonia, and methane ice at extreme pressure condition reaching 8 Mbar [2]. All four giant planets in our solar system presumably grew so large because large amounts of ice beyond the ice lines contributed substantially to the accretion of core material. One may therefore assume that Jupiter and Saturn have large dense cores consisting of differentiated rock and ice components at pressures on the order of 10 Mbar in Saturn [2] and 39–64 Mbar in Jupiter [3]. However, recent *ab initio* calculations suggest that such an ice layer would dissolve into the layer of metallic hydrogen above [4]. This work implies that the cores of Jupiter and Saturn have been at least partially eroded. Therefore, when one determines the size of Saturn's core from Cassini mission data or that of Jupiter's core from the upcoming Juno mission, one may in fact only capture a snapshot in time since either core may have been substantially bigger when the planets formed. Core erosion is also expected to occur on exoplanet, many which are much bigger and hotter.

The behavior of water ice at megabar pressures is therefore of great interest in planetary science. Static diamond anvil cell experiments have reached up to 2.1 megabars [5–8] for ice but have attained higher pressures for other materials. Shock wave experiments [9] have reached higher pressures but since they heat the sample significantly it melts at the highest pressures. Dynamic ramp compression techniques [10, 11] are expected to reach high pressures at comparatively low temperatures in the future.

The phase diagram of water is extremely complex because hydrogen bonding between different wwater molecules allows for many different structural arrangements. Fifteen solid phases have been observed experimentally and three have been predicted theoretically. The three most recently synthesized experimental phases, ice XIII, XIV, and XV [12, 13], are the hydrogen-ordered equivalents of disordered phases V, XII, and VI. All were obtained by doping ice with hydrochlorid acid, which allowed the hydrogen-bonded networks to reach the ordered ground state.

Ice X forms at 0.6 Mbar and is the highest-pressure phase seen in experiments. All knowledge of ice above 2 Mbar relies primarily on *ab initio* simulations. In 1996, Benoit *et al.* [14] predicted ice X to exhibit a phonon instability at approximately 3 Mbar that leads to a new phase with *Pbcm* symmetry. A shift between adjacent atomic layers is accommodated by a doubling of the unit cell from 6 to 12 atoms.

 Most recently, Militzer and Wilson [15] predicted the *Pbcm* phase to exhibit another phonon instability at 7.6 Mbar that leads to structure with *Pbca* symmetry with a 24 atom in the unit cell. As a result of the increasing pressure, the hydrogen atoms are squeezed out of the midpoint between the nearest oxygen atoms.

All ice structures that we discussed so far are insulators. Militzer and Wilson predicted ice to become metallic at 15.5 Mbar when a new phase with *Cmcm* symmetry emerges. This transition is expected to greatly increase reflectivity, which will make it easier to be detected spectroscopically in shock wave experiments.

The discussed series of phase transformations is summarized in figure 1 but we still recommend reviewing article [15] before proceeding with this paper. Here, we compare the bonding and electronic properties between ice X and the new *Cmcm* phase. We have selected ice X rather than the *Pbcm* or *Pbca* phase at higher pressure because it has a small unit cell with only 6 atoms. Also the phonon instabilities that lead to the *Pbcm* and *Pbca* phases do not introduce fundamental structural changes. Ice X, *Pbcm,* and *Pbca* phase share many structural properties, e.g., they are composed of two interpenetrating hydrogen-bonded networks. The *Cmcm* structure is composed of corrugated H-O sheets instead.

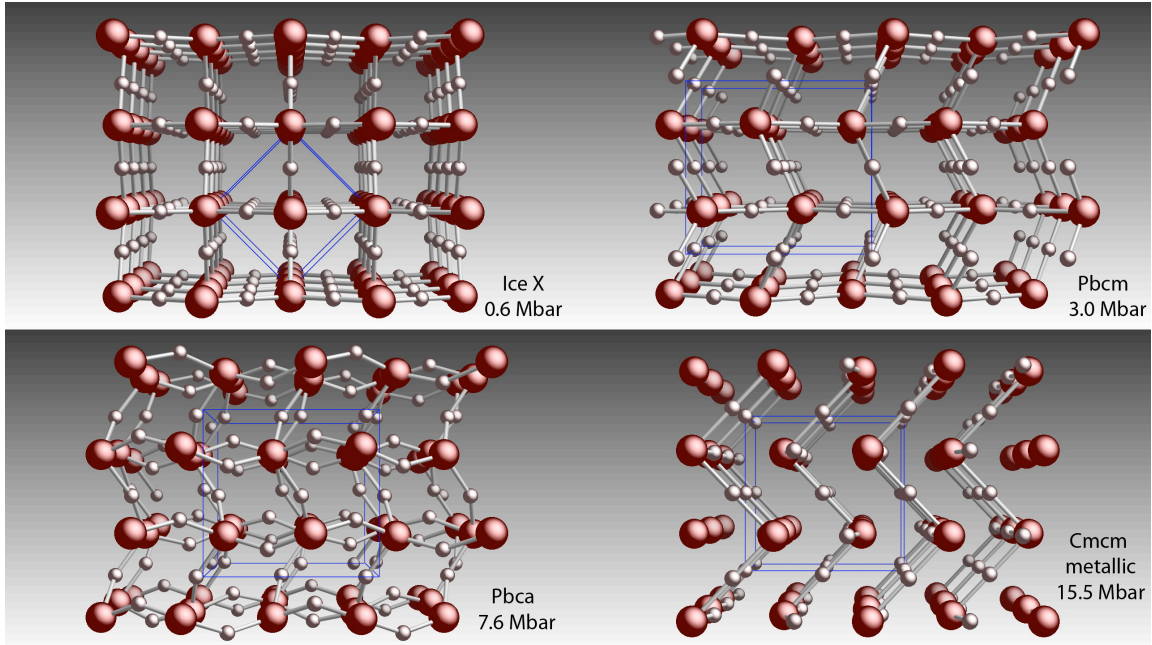

**Figure 1:** Sequence of structural transformations in ice at megabar pressures. Ice X is the highest-pressure phase seen in experiment. The large and small spheres respectively denote the oxygen and hydrogen atoms. The thin lines indicate the unit cells. The *Pbca* and *Cmcm* phases were predicted recently with *ab initio* computer simulations [15]. Similar techniques were used to identify the *Pbcm* phase [14]. The ice X to *Pbcm* transition is caused by a displacement of atomic layers. In the *Pbca* phase, the hydrogen atoms are squeezed out of the midpoint between the nearest oxygen atoms. In the *Cmcm* phase, the hydrogen atoms occupy midpoints between the next-nearest oxygen atoms.

## Methods

We used the Vienna *ab initio* simulation package (VASP) [16, 17] to perform density functional calculations with the generalized gradient approximation by Perdew, Burke, and Ernzerhof [18] in combination with the projector augmented-wave method [19]. We employed the Abinit code [20] to perform the analysis of the Fermi surface. The Wannier orbitals were calculated with the Quantum Espresso [21] and the Wannier90 [22] packages.

## Results

We begin our discussion of the electronic properties of the *Cmcm* phase by analyzing the Fermi surface that is shown in figures 2 and 3 at 12.5 and 22.0 Mbar. As shown in the band structure plot in figure 4 of reference 15, the conduction band dips below the Fermi level at the A point in the monoclinic unit cell. This leads to a round disc-shaped surface that increases in size with pressure. Two valence bands reach above the Fermi level near the Γ point, which leads to two more Fermi surface segments shown in figure 2. The inner segment comes from a band that barely reaches above the Fermi level.

The *Cmcm* structure was found to be metallic for all pressures under consideration. However, in [15], it was demonstrated that for pressures up to 15.5 Mbar, the *Cmcm* structure is subject to a Peierls distortion that opens up a band gap and leads to a structure with *Pnma* symmetry in a 24-atom unit cell. The *Pnma* and *Cmcm* structures show slight differences in lattice parameters and in the atomic positions. Recent calculations revealed that it is solely the shift in atomic positions that is responsible for the opening of the band gap.

The *Cmcm*-to-*Pnma* distortion could be caused by a Fermi surface that is *nested*, i.e., when one large segment of the Fermi surface is parallel to another [23]. This may cause the formation of a charge density wave with a wave vector $q$ that is identical the vector that separates the two Fermi surface segments in k-space. A charge density wave can introduce electronic instabilities and lead to a metal-to-insulator transition. In particular, low-dimensional metals with partially filled bands that are dispersive primarily along one direction are susceptible to such electronic instabilities [24]. In such cases, one would expect to find two *flat* and parallel segments in the Fermi surface.

The Fermi surface plots in figure 2 and 3 exhibit primarily round shapes. Especially, the disc at the A point is not consistent with a low-dimensional conductor. However, the top view in figure 3 reveals the existence of a segment in the conduction band that is separated approximately by the vector q= (½, ½, 0) from a section of the valence band surface. Both are only approximately parallel, which means nesting effects may not be very strong [25] but may nevertheless be related to the observed *Cmcm*-to-*Pnma* instability. The vector q= (½, ½, 0) is perfectly commensurate with the observed structural distortion <u>that leads to the</u> orthorhombic *Pnma* unit cell with 12 atoms. This adds support the hypothesis that Fermi nesting is at least partially responsible for the *Cmcm*-to-*Pnma* instability. On the other hand, this instability is only observed for pressures below 15.5 Mbar where the Fermi surface is very small and it is difficult to identify the parallel segments. In conclusion, two parallel nearly parallel Fermi surface segments have been identified but further research is needed to understand all implications.

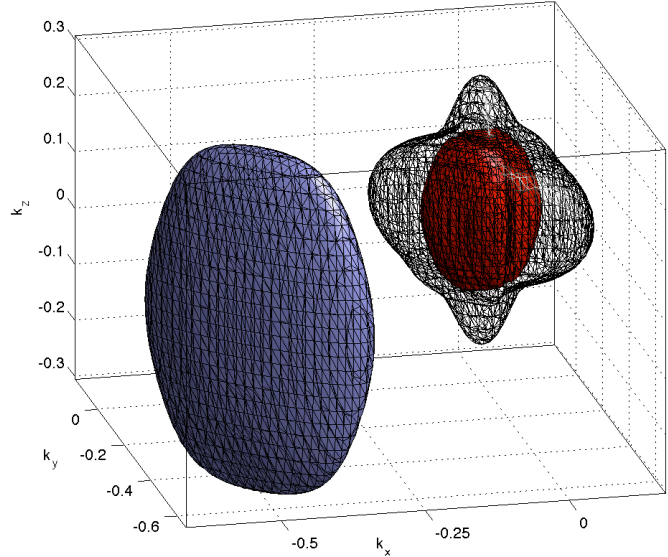

**Figure 2:** Fermi surface of the *Cmcm* phase at 22 Mbar is plotted in the reciprocal space of the 6-atom monoclinic unit cell. The large disc on the left is caused by the conduction band dipping below the Fermi level at the A point, k=(½, ½, 0), of the Brillouin zone. The other two surfaces, centered at the Γ point, are the result of two valence bands reaching above the Fermi level.

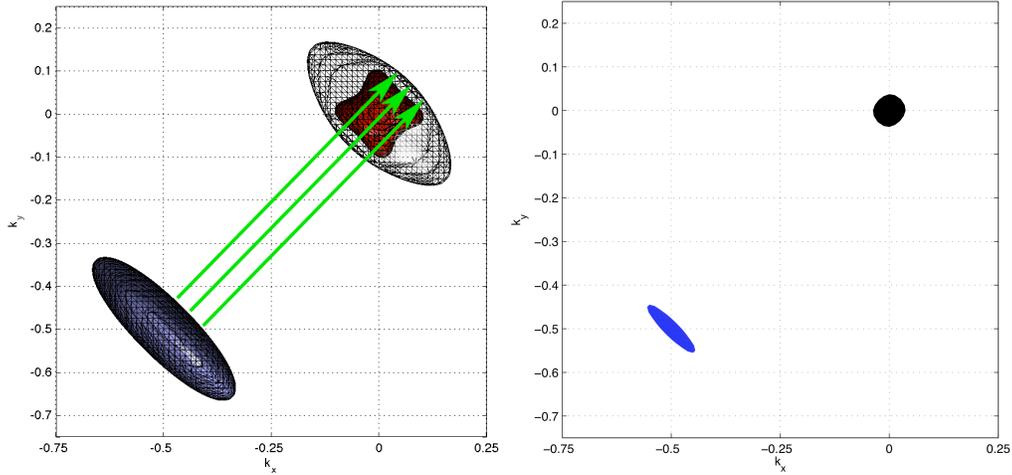

**Figure 3:** Fermi surfaces of the *Cmcm* phase at 22.0 and 12.5 Mbar are show on the left and right, respectively. The left diagram shows the top view of figure 2. The arrows indicate two nearly parallel Fermi surface segments that are separated by approximately the vector *q*=(½, ½, 0).

In this section, we compare the bonding properties in ice X and *Pnma* structure by computing maximally localized Wannier orbitals. Since this analysis is only applicable to insulators, we applied it to the *Pnma* structure rather than to the *Cmcm* phase. However, the total charge densities of both structures are very similar, so that at least some conclusions may also apply to the *Cmcm* structure. While the ice X, *Pbcm*, and *Pbca* structures have two interpenetrating hydrogen bonded networks, the *Pnma* structure is composed of corrugated O-H sheets instead. The main goal of the Wannier analysis is to see whether this leads to any changes in the bonding pattern.

We expect to find four Wannier orbitals per water molecule to be filled by six valence electrons from the oxygen atom and two electrons from the hydrogen atoms. Figure 4 reveals that ice X is a classic example of $sp^3$ hybridization as found in, e.g., methane or diamond. The main lobe of each Wannier orbital is aligned with a hydrogen bond. The minor lobe is found on the opposite side of the oxygen atom. Each hydrogen atom is touched by two Wannier orbitals, one from each nearby oxygen atom. Most of the electronic charge is localized near the oxygen atom but some flows outwards along the hydrogen bonds.

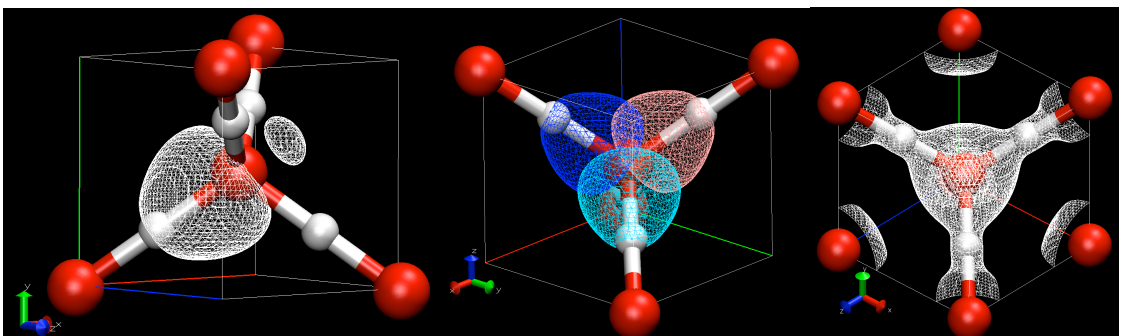

**Figure 4:** Wannier orbitals and charge density of ice X at 3 Mbar. The left panel illustrates the $sp^3$ hybridization showing just one Wannier orbital with its main lobe oriented towards the hydrogen atom and its minor lobe on the opposite side. The center image shows the orientation of three Wannier orbitals in different color that bind the central oxygen atom to nearby hydrogen atoms. (For clarity a higher isosurface level was chosen so that only the main orbital lobes appear. Also the fourth Wannier orbital that binds the oxygen to the hydrogen atom on the back side was omitted.) The isosurface plot on the right shows the total charge density.

Figures 5 and 6 demonstrate that a different picture emerges when the same analysis performed for the *Pnma* structure. While it is still true that the total electronic charge is located near the oxygen atoms and along the hydrogen bonds (figure 6), the Wannier orbitals have changed. Most strikingly, one of the four orbitals is no longer aligned with a particular hydrogen bond. Instead, it sticks out in Z direction and has a much rounder shape than the three others. Those three have also changed shape slightly compare to ice X. They are no longer radially symmetric around the axis of the hydrogen bond.

The hydrogen atoms in the *Pnma structure* are no longer equivalent. The hydrogen atom that links two oxygen atoms of the same sheet has only one Wannier orbital pointing in its direction, while all remaining hydrogen atoms have two, as all hydrogen atoms in the ice X structure have. The plot of total charge density in figure 6, however, reveals that there is plenty of electronic charge localized along bonds of this atom with its two oxygen neighbors. This is again an indication that the Wannier orbitals have adopted a different shape compared to ice X structure that allows them to donate charge in the direction of the hydrogen atoms while their core is pointing in a different direction.

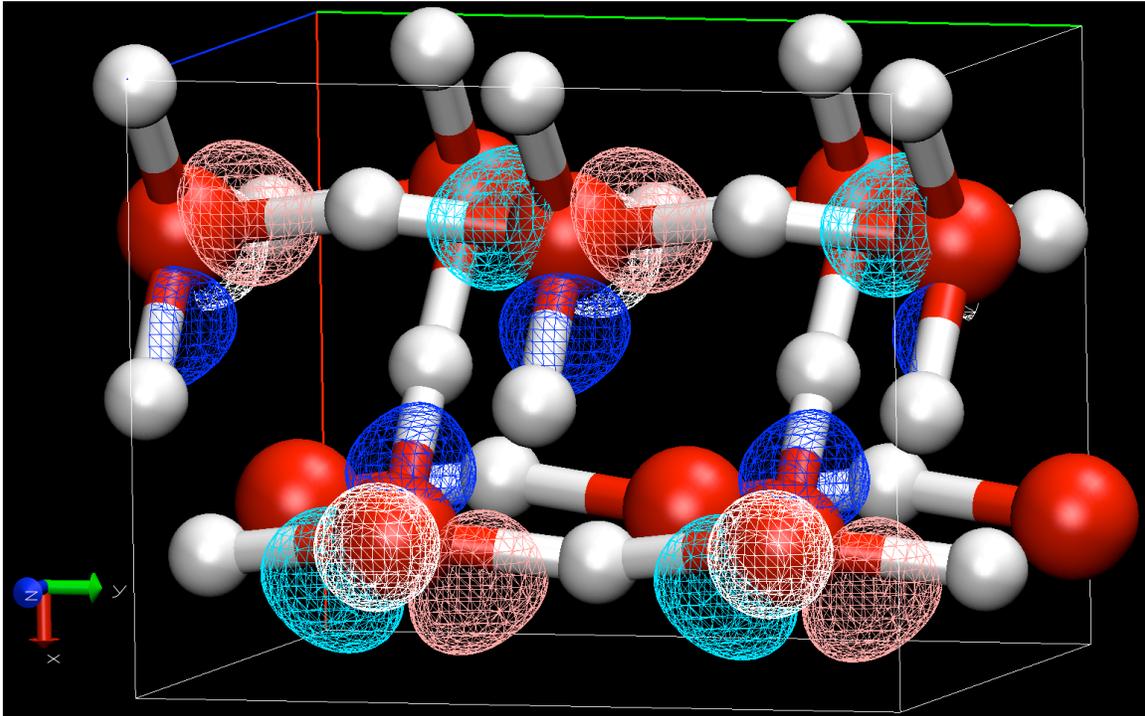

**Figure 5:** Wannier orbitals in the *Pnma* phase at 12.5 Mbar are shown in different colors. Only orbitals of the atoms in front are shown. While the three orbitals shown in darker colors are all aligned with a particular hydrogen bond, the white orbital that is oriented in Z direction shows no such alignment. Conversely, in Ice X, all orbitals are aligned (figure 4).

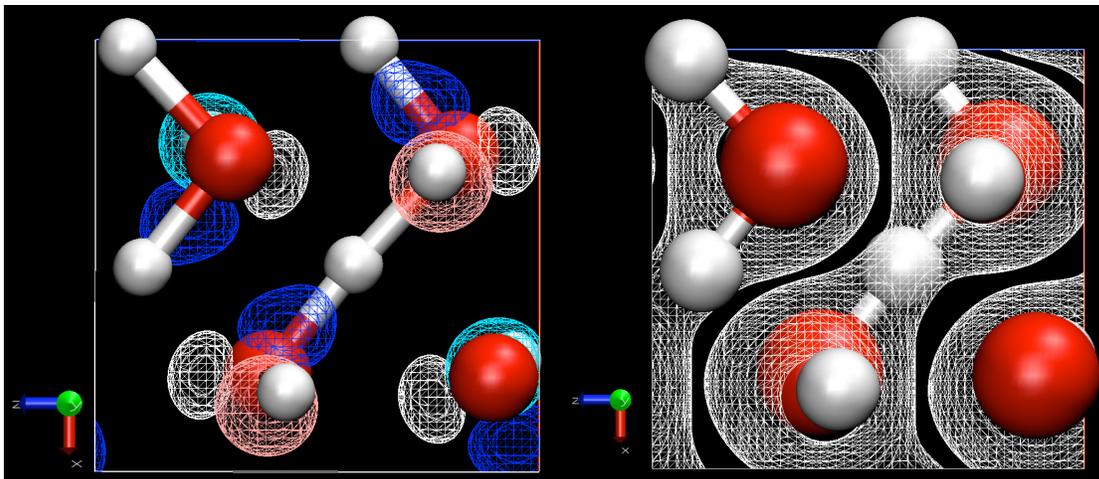

**Figure 6:** Left: side view of the Wannier orbitals in the *Pnma* phase at 12.5 Mbar from figure 5. The right image shows the corresponding total charge density. The cell orientation is similar to the *Cmcm* structure in figure 1. The O-H sheets are parallel to the XY plane and corrugated into the Z direction. The hydrogen atom in the center of the image links two oxygen atoms of the same sheet.

While the directions of the hydrogen bonds in *Pnma* have changed significantly compare to ice X, the Wannier orbitals have not fully adopted that change. They do not deviate very much from the typical $sp^3$ orientations. The resulting misalignment of Wannier orbitals and the hydrogen bonds most likely leads to an unfavorable increase in internal energy, E. However, the *Pnma* structure becomes stable only at megabar pressures when its denser packing becomes sufficiently important so that its low PV term outweighs such an energy increase.

## Conclusions

We have discussed the properties of water ice at megabar pressures and focused on the recently predicted metallic *Cmcm* structure and the slightly distrorted version with *Pnma* symmetry. An analysis of *Cmcm* Fermi surface showed some evidence of Fermi nesting but more work will be needed to understand this process in more detail and its implications at yet higher pressures.

Our analysis of the Wannier orbitals revealed substantial differences in the bonding properties between ice X and the *Pnma* structure. Ice X is a typical $sp^3$ bonded material where the Wannier orbitals are aligned with the hydrogen bonds. In the *Pnma* structure, the shape of the Wannier orbitals is different and one out of four is not aligned with any hydrogen bonds.

## Acknowledgements

The author acknowledges support from NSF, UC's lab fee program, and NASA. Computational resources at NCCS, NERSC, and TAC were used. We thank R. Caracas for suggesting the Fermi nesting analysis.